\documentclass[prb,twocolumn,showpacs,amsmath,amssymb,floats]{revtex4}
\usepackage[dvips]{graphicx}

\def\be{\begin{equation}}
\def\ee{\end{equation}}
\def\bea{\begin{eqnarray}}
\def\eea{\end{eqnarray}}

\begin{document}

\title{The gap exponent of XXZ model in a transverse field}

\author{A. Langari$^{1,2}$ and S. Mahdavifar$^3$}
\affiliation{$^1$Physics Department , Sharif University of Technology, Tehran 11365-9161, Iran\\
$^2$Institute for Studies in Theoretical Physics and Mathematics (IPM), Tehran 19395-5531, Iran\\
$^3$Institute for Advanced Studies in Basic Sciences, Zanjan  45195-1159, Iran
}
\date{\today}
%\pacs{75.10.Jm}{Quantized spin models}
%\pacs{75.10.Pq}{Spin chain models}
%\pacs{75.40.Cx}{Static properties (order parameter, static susceptibility,
%heat capacities, critical exponents, etc) }

\begin{abstract}
We have calculated numerically the gap exponent of the anisotropic
Heisenberg model in the presence of the transverse magnetic field.
We have implemented the modified Lanczos method to obtain the
excited states of our model with the same accuracy of the ground
state. The coefficient of the leading term in the perturbation expansion
diverges in the thermodynamic limit ($N\rightarrow \infty$). We have obtained the
relation between this divergence and the scaling behaviour of the energy gap.
We have found that the opening of gap in the presence of
transverse field scales with a critical exponent which depends on
the anisotropy parameter ($\Delta$). Our numerical results
are in well agreement with the field theoretical approach in the  whole range of
the anisotropy parameter $-1 < \Delta < 1$.

\end{abstract}
\maketitle

\section{introduction}
The study of transverse magnetic field on the low dimensional spin
systems has been attracted much interest recently from
experimental and theoretical point of view. The experimental
observations \cite{kenzelmann, cpfs} on the quasi-one-dimensional
spin-$1/2$ anti-ferromagnet $Cs_2CoCl_4$ is a realization of the
effect of non-commuting field on the low energy behavior of a
quantum model. This shows a quantum phase transition from the
spin-flop phase (ordered antiferromagnetically in y-direction) at
the low magnetic filed to a paramagnet for high fields. Moreover a
novel behavior has been observed in the specific heat close to the
quantum critical point \cite{cpfs}. A connection between the
ground state properties of the anisotropic Heisenberg model (XXZ)
in the transverse field to the reported quantum phase transition
has been given by the quantum renormalization group approach
\cite{langari-qrg}. In addition, a recent mean field approach for
the weakly coupled chains \cite{dima3} has given a very good
agreement on the finite temperature phase diagram of the two
dimensional model with the observed experimental data
\cite{kenzelmann}.

The spin-$(s=\frac{1}{2})$ Hamiltonian of the XXZ model in a
transverse field on a periodic chain of N-sites is
%%%%%%%%%%%%%%%%%%%%%%%%%%%%%%%%%%%%%%%%%%
\begin{equation}
H=J\sum_{i=1}^N [s^x_i s^x_{i+1}+ s^y_i s^y_{i+1} +\Delta
s^z_is^z_{i+1} -h s^x_i],
\label{ham}
\end{equation}
%%%%%%%%%%%%%%%%%%%%%%%%%%%%%%%%%%%%%%%%%%
where $J>0$ is the exchange couplings in the XY easy plane, $-1
\leq \Delta < 1$ is the anisotropy in Z direction and $h$ is
proportional to the transverse field. The spin ($s^{\alpha}_n$) on
site $n$ is represented by $\frac{1}{2}\sigma^{\alpha}_n;
\alpha=x, y, z$  in terms of Pauli matrices. This model is a good
candidate for explaining the low temperature behavior of
$Cs_2CoCl_4$. However, since the integrability of XXZ model will
be lost in the presence of transverse filed, more intensive
studies from the theoretical point of view is invoked.

When $h=0$, the XXZ model is known to be solvable and
critical (gapless)~\cite{yang}. The Ising regime is governed by $\Delta > 1$
while for $\Delta \leq -1$ it is in the ferromagnetic phase.
Magnetic field in the anisotropy direction commutes with
the Hamiltonian at $h=0$ and extends the gapless region (quasi long range
order) to a border where a transition to paramagnetic phase takes place.
The model is still integrable and can be explained by
a conformal field theory with central charge $c=1$ (Ref.[\onlinecite{alcaraz}] and references therein).

Adding a transverse field to the XXZ model breaks the $U(1)$
symmetry of the Hamiltonian to a lower, Ising-like, which develops
a gap. The ground state then has long range anti-ferromagnetic
order ($-1 < \Delta < 1$). However due to non-zero projection of
the order parameter on the field axis it is a spin-flop N\'{e}el
state. In fact at a special field ($h_{cl}=\sqrt{2(1+\Delta)}$) the
ground state is known exactly to be of the classical N\'{e}el type
~\cite{muller,mori1}. The gap vanishes at the critical field
$h_c$, where the transition to the paramagnetic phase occurs.
Classical approach to this model reveals the mean field results
~\cite{kurmann} which is exact at $s\rightarrow \infty$. The
implementation of the quantum renormalization group
\cite{langari-qrg} shows that the transition at $h_c$ is in the
universality class of the Ising model in Transverse Field (ITF).
The phase diagram of XYZ model in transverse field has been also
presented in Ref.[\onlinecite{langari-qrg}]. The scaling of gap,
phase diagram and some of the low excited states at $h_{cl}$ of the
XXZ model in the transverse field has been studied in
Ref.[\onlinecite{dmitriev}]. In this approach the scaling of gap
is given by the scaling of operator $s^x$ which is read from the
asymptotic form of the correlation function \cite{luther}
($\langle s^x_i s^x_{i+r} \rangle$). This correlation function
contains two terms, an oscillating and a non-oscillating one. Each
part defines a specific scaling exponent for  $s^x$ which depends
on the anisotropy parameter ($\Delta$).

Exact diagonalization \cite{mori} and Density Matrix
Renormalization Group (DMRG) \cite{capraro} results give us some
knowledge on this model but not on the scaling of gap. A
bosonization approach to this model in certain limits leads to a
nontrivial fixed point and a gapless line which separates two
gapped phases \cite{dutta}, moreover the connection to the axial
next-nearest neighbor Ising model (ANNNI) has been addressed. The
applicability of the mean-field approximation has been studied by
comparing with the DMRG results of magnetization and structure
factor\cite{caux}. Recently the effect of longitudinal magnetic
field on both the Ising model in Transverse Field (TF)
\cite{ovchinnikov} and the XXZ model in TF has been discussed
\cite{dima2}.

Here, we are going to present our numerical results on the low
energy states of the XXZ model in the transverse field which has been
obtained by the modified Lanczos method introduced in Sec.II. It
is believed that in this approach the accuracy of the excited
state energies is the same as the ground state energy.  
In Sec.III, we will apply a scaling argument presented in
Ref.[\onlinecite{fledderjohann}] to our model and examine it by
the numerical results to obtain the gap exponent.
We have then applied a perturbative approach in Sec. IV to study the
divergent behavior of the coefficient in the  leading term of the perturbation expansion.
The gap exponent has been obtained by the relation to the 
exponent of the diverging term. The gap exponent depends
on the anisotropy parameter in agreement with the field theoretical results \cite{dmitriev}. 
Finally, we will present the summary and discussion on our
results.

%%%%%%%%%%%%%%%%%%%%%%%%%%%%%%%%%%%   section

\section{Modified Lanczos method}
The theoretical investigation of numerous physical problems
requires an appropriate handling of matrices of very large rank.
Even if in many applications the matrix is sparse, the problem can
not be solved by means of a direct diagonalization by standard
routines. The Lanczos method and the related recursion methods
\cite{lanczos,haydock,grosso,lin}, possibly with appropriate
implementations, have emerged as one of the most important
computational procedures, mainly when a few extreme eigenvalues
(largest or smallest) are desired. Grosso and Martinelli have
presented a relevant implementation of the Lanczos
tri-diagonalization scheme \cite{ggrosso}, which allows to obtain a
very fast convergence to any excited eigenvalue and eigenfunction
of H, overcoming memory storage difficulties. To explain this
method briefly, let us consider an operator H, with unknown
eigenvalues $E_{i}$ and eigenfunctions $|\psi_{i}\rangle$. Any
auxiliary operator $A=f(H)$ commutes with H, and thus shares with
it a complete set of eigenfunctions corresponding to the
eigenvalues $A_{i}=f(E_{i})$. In order to obtain the nearest excited state
of H to any apriori chosen trial energy
$E_{t}$, we consider the auxiliary operator A in the form
$A=(H-E_{t})^2$. In a completely different context, this form is
suggested by numerical analysis \cite{nash} to solve the
schrodinger equation within any desired energy. Now, we are faced
with the solution of the following eigenvalue equation
%%%%%%%%%%%%%%%%%%%%%%%%%%%%%%%%%%%%%%%%%%
\begin{equation}
A|\psi_{i}\rangle\equiv(H-E_{t})^{2}\mid\psi_{i}\rangle=\lambda_{i}|\psi_{i}\rangle
\label{gros}
\end{equation}
%%%%%%%%%%%%%%%%%%%%%%%%%%%%%%%%%%%%%%%%%%
Our strategy to solve Eq.($\ref{gros}$) is based on the Lanczos
algorithm. We briefly summarize some basic features of the Lanczos
procedure in its standard formulation.\\
Let us denote with ${\varphi_{i}}$ $(i=1,2,...,N)$ a complete set
of bases functions, for the representation of the operator H (and
hence of A). Starting from a seed state $|u_{0}\rangle$, given by
whatever chosen linear combination of the ${\varphi_{i}}$, a set
of orthonormal states
$|u_{0}\rangle|u_{1}\rangle,...,|u_{N}\rangle$ is constructed via
successive applications of the operator A as follows
\bea
|U_{1}\rangle=(A-a_{0})|u_{0}\rangle, \nonumber \\
a_{0}=\langle u_{0}|A|u_{0}\rangle.
\eea
In general
%%%%%%%%%%%%%%%%%%%%%%%%%%%%%%%%%%%%%%%%%%
\begin{equation}
|U_{n+1}\rangle=A|u_{n}\rangle-a_{n}|u_{n}\rangle-b_{n}|u_{n-1}\rangle,~~~~n>1
\label{unp1}.
\end{equation}
%%%%%%%%%%%%%%%%%%%%%%%%%%%%%%%%%%%%%%%%%%
The (non-normalized) state $|U_{n+1}\rangle$ allows us to
determine the coefficients $b_{n+1}$ and $a_{n+1}$ of the (n+1)th
iteration step, via the procedure
%%%%%%%%%%%%%%%%%%%%%%%%%%%%%%%%%%%%%%%%%%
\begin{eqnarray}
b_{n+1}^{2}=\langle U_{n+1}|U_{n+1}\rangle, \nonumber  \\
a_{n+1}=\frac{\langle U_{n+1}|A|U_{n+1}\rangle}{\langle
U_{n+1}|U_{n+1}\rangle},
\label{coefficients}
\end{eqnarray}
%%%%%%%%%%%%%%%%%%%%%%%%%%%%%%%%%%%%%%%%%%
where $b_0=0$ is the initial condition. After normalization of the
state $|u_{n+1}\rangle=\frac{1}{b_{n+1}}|U_{n+1}\rangle$, the
steps ($\ref{unp1}$) and ($\ref{coefficients}$) are repeated with
n replaced by n+1. In the new basis ${|u_{n}\rangle}$, the
operator A is represented by a tridiagonal matrix $T_{m}$, whose
elements $a_{n}$ and $b_{n}$ are explicitly known for $( m\leq N
)$. The diagonalization of tridiagonal matrix $T_{m}$ gives the
eigenvalues, $\lambda_{i}=(E_{i}-E_{t})^2$.

The transformation to the tridiagonal matrix is truncated at some
stages because of the round-off error. However, the ground state
energy can be obtained up to some significant digits. The accuracy
of the excited energies is lost in the usual Lanczos method
($A=H$) by the round-off error. The modified Lanczos method explained
above allow us to get the higher energy levels with the same
accuracy of the ground state energy. We can select the tuning
parameter $E_t$ in the range of accuracy of our method to get the
excited energy levels. By choosing the appropriate $E_t$, we got
the energy gap of the model presented in Eq.($\ref{ham}$) up to 8
digits. 
%The results for some interval of the transverse field has been
%presented in Table I. The energy gap is defined by the difference
%energy between the second excited  and the ground state
%($gap=E_2-E_0$) which will be discussed in next section.

%%%%%%%%%%%%%%%%%%%%%%%%%%%%%%%%%%%   section

\section{The scaling argument and gap exponent}

The XXZ model is integrable and its low-energy properties are
described by a free massless boson field theory. In the transverse
magnetic field, Dimitriev et.al considered \cite{dmitriev},
the perturbed action for the model as
%%%%%%%%%%%%%%%%%%%%%%%%%%%%%%%%%%%%%%%%%%
\begin{equation}
S=S_{0}+h \int dt dx S^{x} (x,t),
\end{equation}
%%%%%%%%%%%%%%%%%%%%%%%%%%%%%%%%%%%%%%%%%%
where $S_{0}$ is the Gaussian action of the XXZ model. The time
dependent correlation function of the XXZ chain for
$\mid\Delta\mid<1$ has the following asymptotic form \cite{luther}
%%%%%%%%%%%%%%%%%%%%%%%%%%%%%%%%%%%%%%%%%%
\begin{eqnarray}
\nonumber
\langle S^{x}(x,\tau) S^{x}(0,0) \rangle&&\sim\frac{(-1)^{x}
A_{1}}{(x^{2}+v^{2}
\tau^{2})^{\theta/2}}\\
&&-\frac{A_{2}}{(x^{2}+v^{2}
\tau^{2})^{(\theta/2+1/2\theta)}},
\label{correlation}
\end{eqnarray}
%%%%%%%%%%%%%%%%%%%%%%%%%%%%%%%%%%%%%%%%%%
where $A_{1}$ and $A_{2}$ are known constants \cite{hikihara},
$\tau=it$ is the imaginary time, $v$ is the velocity and
%%%%%%%%%%%%%%%%%%%%%%%%%%%%%%%%%%%%%%%%%%
\begin{equation}
\theta=1-\frac{arc\cos\Delta}{\pi}.
\end{equation}
%%%%%%%%%%%%%%%%%%%%%%%%%%%%%%%%%%%%%%%%%%
The exponent of energy gap has been estimated \cite{dmitriev} by
using the long-distance contribution of the oscillating part to
the action via the scaling of $S^{x}$
%%%%%%%%%%%%%%%%%%%%%%%%%%%%%%%%%%%%%%%%%%
\begin{equation}
gap\sim h^{\mu},~~~~~~~~~~~
\mu=\frac{1}{1-\theta/2}.
\label{mu}
\end{equation}
%%%%%%%%%%%%%%%%%%%%%%%%%%%%%%%%%%%%%%%%%%
If the scaling of $S^{x}$ is read from the non-oscillating part of
$\langle S^{x}(x,\tau) S^{x}(0,0) \rangle$ the contribution to the
action gives the following scaling for the energy gap
\cite{gogolin,nersesyan}
%%%%%%%%%%%%%%%%%%%%%%%%%%%%%%%%%%%%%%%%%%
\begin{equation}
gap\sim h^{\nu},~~~~~~~~~~~~\nu=\frac{2}{4-\theta-1/\theta}.
\label{nu}
\end{equation}
%%%%%%%%%%%%%%%%%%%%%%%%%%%%%%%%%%%%%%%%%%
The smaller value of $\mu$ and $\nu$ defines the leading order of
the dependence of gap on the transverse field, the {\it gap exponent}.
Thus, $\mu$ is the gap exponent for $-1 < \Delta \le 0$ and $\nu$
for $0\le \Delta < 1$.
 
%%%%%%%%%%%%%%%%%%%%%%%%%%%%%%%%%

%%%%%%%%%%%%%%%%%%%%%%%%%%%%%%%%%

%%%%%%%%%%%%%%%%%%%%%%%%%%%%%%%%%%%%%%%%%
\begin{figure}
\centerline{\includegraphics[width=7cm,angle=0]{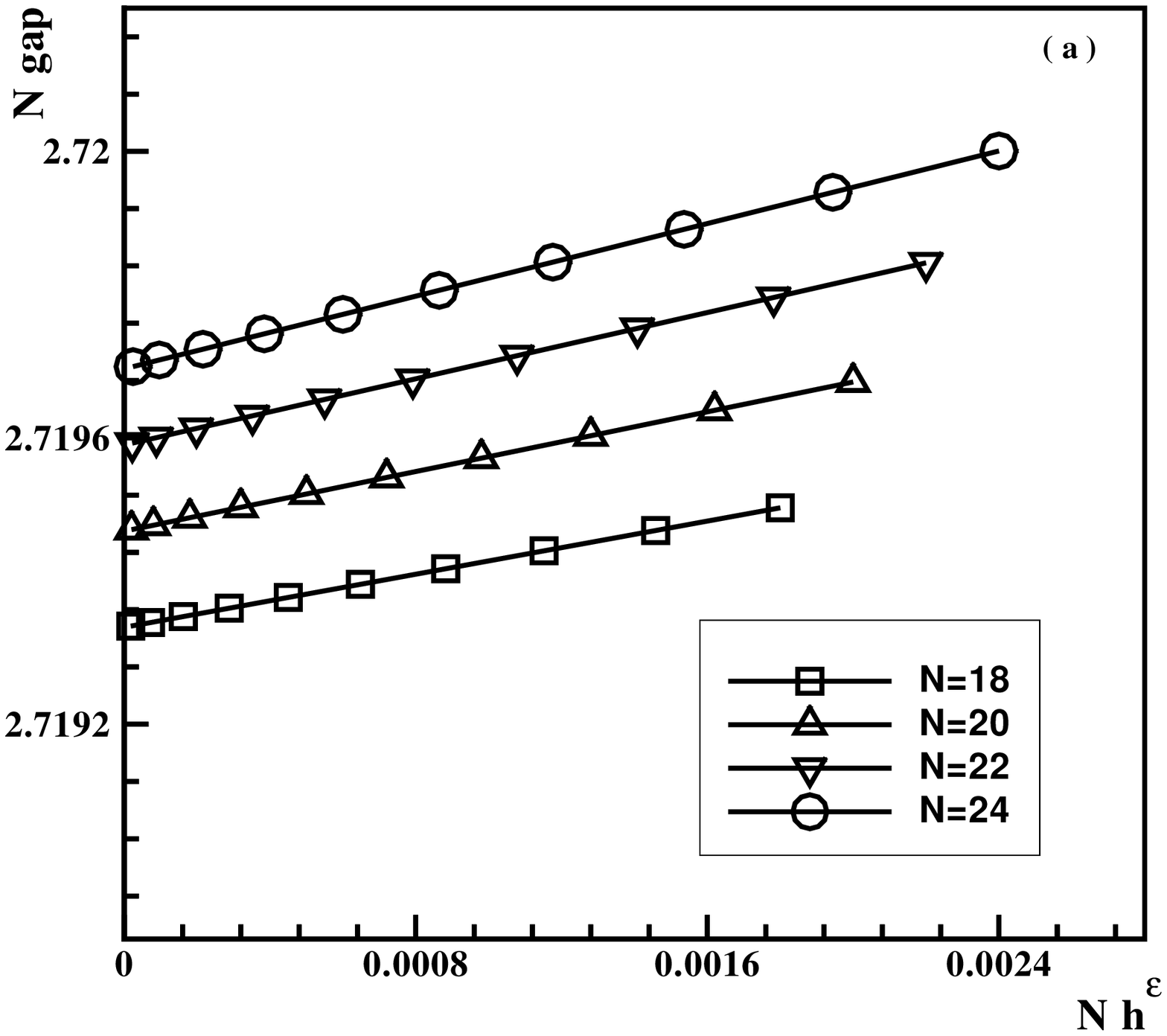}}
\centerline{\includegraphics[width=7cm,angle=0]{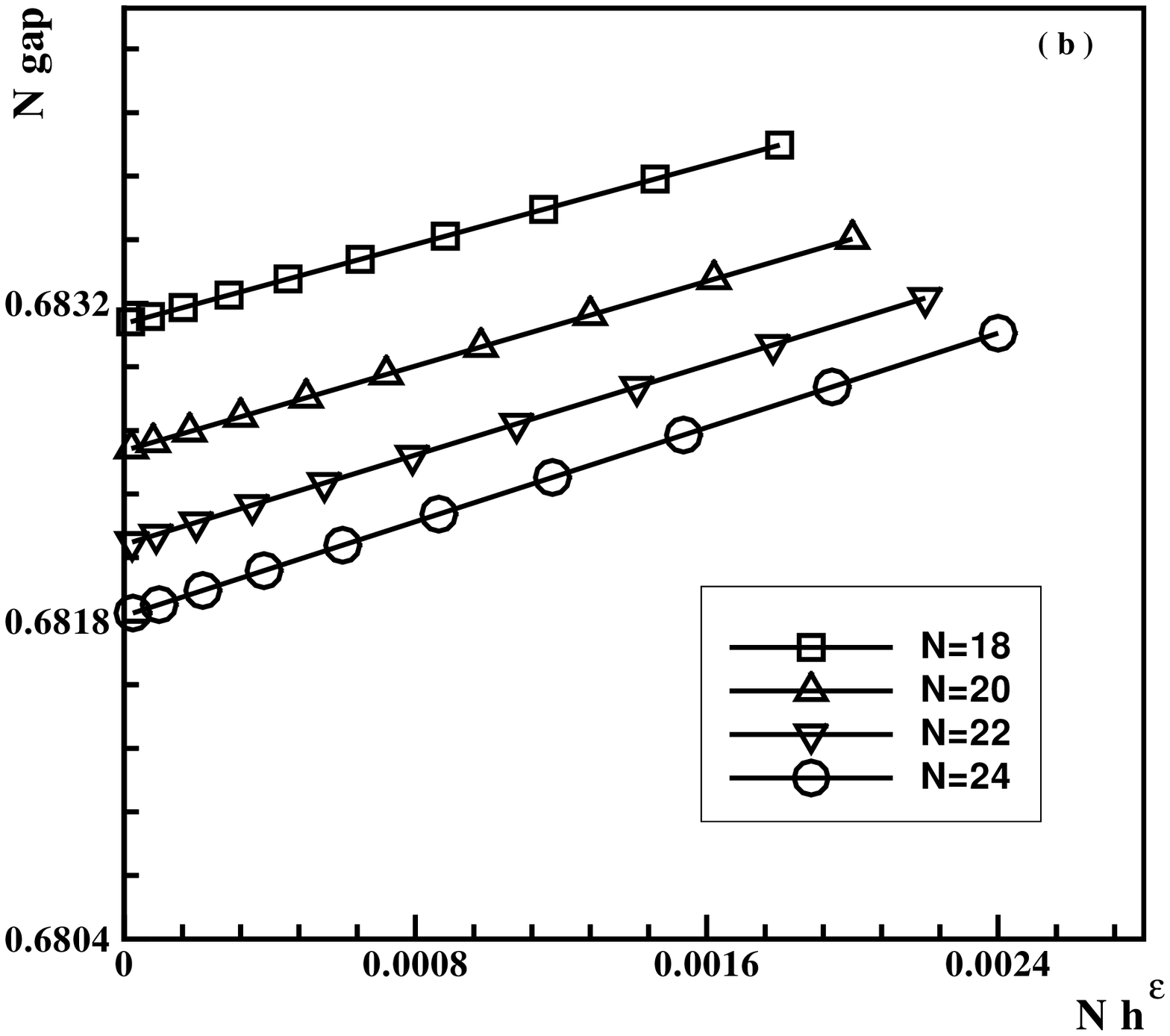}}
\caption{The product of energy gap and chain length ($N gap$) versus
$N h^{\varepsilon}$. (a) $\Delta=0.5$ (b) $\Delta=-0.5$. 
A linear behavior is obtained by choosing $\varepsilon=2.0$
for all different chain lengths $N=18, 20, 22, 24$. The solid lines are guides for eye.
} \label{fig1}
\end{figure}
%%%%%%%%%%%%%%%%%%%%%%%%%%%%%%%%%%%%%%%%%%
We have implemented the modified Lanczos algorithm on the finite size
chains (N=8,10,12,...,24) by using periodic boundary conditions to
calculate the energy gap. We have computed the energy gap for
different values of $-1 < \Delta < 1$ and the chain lengths.
The energy gap as a function of the chain length ($N$) and the
transverse field ($h$) is defined
%%%%%%%%%%%%%%%%%%%%%%%%%%%%%%%%%%%%%%%%%%
\begin{equation}
gap(N,h)=E_{2}(N,h)-E_{0}(N,h),
\end{equation}
%%%%%%%%%%%%%%%%%%%%%%%%%%%%%%%%%%%%%%%%%%
where $E_{0}$ is the ground state energy and $E_{2}$ is the second
excited state one. The first excited state crosses the ground
state $N/2$ times for a finite chain and the last crossing occurs
at the classical point\cite{dmitriev} $h_{cl}=\sqrt{2(1+\Delta)}$. These two
states form a twofold degenerate ground state in the thermodynamic
limit where $E_{1}-E_{0}$ vanishes.

In the case of $h$ equals to zero, the spectrum of the XXZ model is
gapless. The gap vanishes in the thermodynamic
limit proportional to the inverse of chain length,
\begin{equation}
\lim_{N\rightarrow\infty}gap(N, h=0){\longrightarrow} \frac{B}{N}.
\label{gaf}
\end{equation}
The coefficient $B$  is known exactly from the Bethe ansatz solution \cite{klumper}.
We consider this equation as the initial
condition for our procedure\cite{fledderjohann}. 
Adding the transverse field to the Hamiltonian, a nonzero gap develops. 
The presence of gap can be characterized by the following expression,
\begin{equation}
\frac{gap(N,h)}{gap(N,0)}=1+f(x),
\label{gaf1}
\end{equation}
in the combined limit
\begin{equation}
N\rightarrow\infty,~~~~h\rightarrow0,
\end{equation}
where $x=N h^{\varepsilon}$ is fixed and $f(x)$ is the scaling function.
Thus, the gap at finite $h$ can be defined
\begin{equation}
gap(N, h)=\frac{B}{N}+\frac{B}{N}f(x)\equiv \frac{B}{N}+ g(N) h^{\varepsilon},
\label{gap}
\end{equation}
where $g(N)$ is a function of only $N$.
It is imposed that the function $g(N)$ approaches
a non-zero constant value in the thermodynamic limit: $\lim_{N\rightarrow \infty} g(N) = constant \equiv C_{\infty}$.
The regime where we can observe the scaling of gap is in the thermodynamic limit ($N\rightarrow \infty$)
and very small value of $h$ ($h\ll 1$). This means that the scaling behavior is observable at large $x$ 
($x=N h^{\varepsilon} \gg 1$). The asymptotic behavior of $g(N)$ for $N\rightarrow \infty$ defines that
the large $x$ behavior of $f(x)$ must be proportional to $x$,
\begin{equation}
f(x)=\frac{g(N)}{B} x \;\;\;\Rightarrow \lim_{x \gg 1} f(x)= \frac{C_{\infty}}{B} x.
\label{largexfx}
\end{equation}
Thus, if we consider the asymptotic behavior of $f(x)$ as 
\begin{equation}
f(x) \sim x^{\phi},
\label{phi}
\end{equation}
the $\phi$-exponent must be 
equal to one ($\phi=1$).
If we multiply both sides of Eq.(\ref{gap}) by $N$ we get,
\begin{equation}
\lim_{N \to \infty (x \gg 1)} N gap(N, h) = B + C_{\infty} x.
\label{ngap}
\end{equation}
Eq.(\ref{ngap}) shows that the large $x$ behavior of $N gap(N, h)$ is linear in $x$ where the
scaling exponent of the energy gap is $\varepsilon$. 

We have plotted the values of $N
gap(N, h)$ versus $N h^{\varepsilon}$ for
$\Delta=0.5,-0.5$ in Fig.(\ref{fig1}). The reported results have
been computed on a chain of length $N=18, 20, 22, 24$ with
periodic boundary conditions.  According to
Eq.(\ref{ngap}), it can be seen from our numerical results
presented in Fig.(\ref{fig1}-a) and Fig.(\ref{fig1}-b) that the
linear beahavior  is very well satisfied  by $\varepsilon=2.0$. 
But, the data do not show a scale invariance plot for different $N$
which is expected from a scaling behavior. We
have also implemented our numerical tool to calculate the exponent
of energy gap  at $\Delta=0, 0.25, -0.25$. Again, the plot of 
$N gap(N, h)$ versus $N h^{\varepsilon}$ shows a linear behavior for
$\varepsilon=2.0$. 
%We have also
%investigated the asymptotic behavior of the scaling function $f(x)$
%with our numerical results. We have found that  the scaling 
%function $f(x)$ is very well parameterized by
%Eq.(\ref{phi}) with $\phi=1.00(4)$.

However, we should note that the horizontal axes presented in Fig.(\ref{fig1})
are limited to very small values of $x=N h^{\varepsilon} < 0.0024$.
Thus, we are not allowed to read the scaling exponent of gap which 
exists in the thermodynamic limit ($N\rightarrow \infty$ or  $x\gg 1$). 
%The plots
%in Fig.(\ref{fig1}) show the small $x$ behavior of the scaling function, $f(x) \sim x$.
We have been limited to consider the maximum value of 
$N=24$. Because for the present model (Eq.(\ref{ham})) the total
$S^z$ does not commute with the Hamiltonian and we should
consider the full Hilbert space of $2^N$ in our computations. Moreover,
to avoid the effect of level crossings, we should  consider very small
values of $h <0.01$. Therefore, the value of $x$ can not be increased 
in this method. We will face up with the same problem even though 
the calculation is done by Density Matrix Renormalization Group (DMRG).
In that case we can extend the calculation for larger sizes, $N \sim 100$ but the 
first level crossing happens for smaller value of $h$. The position of the first level 
crossing is proportional to $N^{-1/\varepsilon}$ which will be explained in the next sections.
This is the level crossing between the higher excited states, in this case between the second and
third ones.
Thus, we have to find  the scaling behavior from the small $x$ data.

%%%%%%%%%%%%%%%%%%%%%%%%%%%%%%%%%%%%%%%%%
\begin{figure}
\centerline{\includegraphics[width=7cm,angle=0]{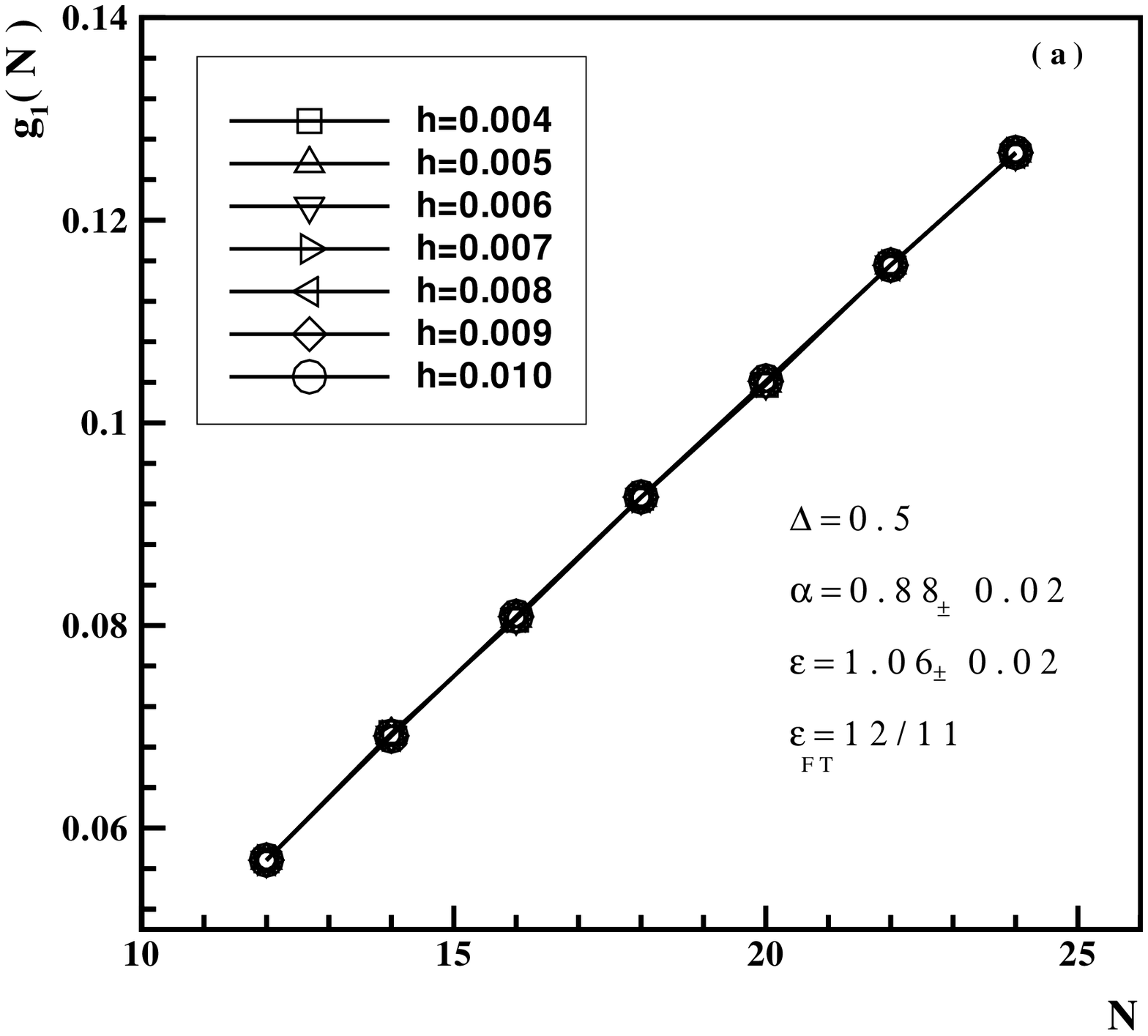}}
\centerline{\includegraphics[width=7cm,angle=0]{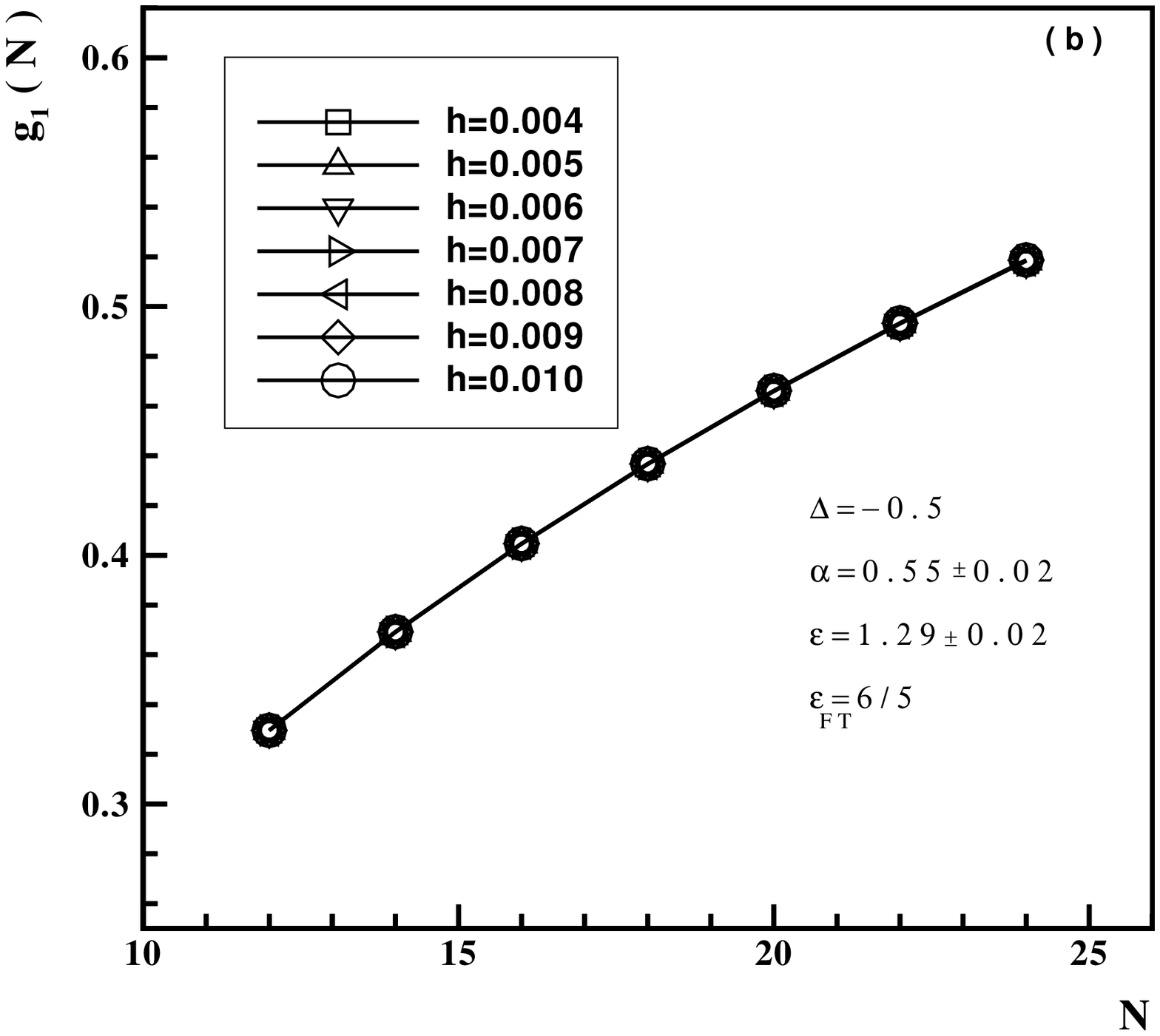}}
\caption{The value of $g_{1}(N)$ versus the chain length ($N$). 
(a) $\Delta=0.5$, (b) $\Delta=-0.5$. Data for different transverse 
fields $0.004\le h\le 0.01$ fall exactly on each other.}
\label{fig2}
\end{figure}
%%%%%%%%%%%%%%%%%%%%%%%%%%%%%%%%%%%%%%%%%%

\section{Perturbative approach}
According to our computations where $N\le 24$, the small $x$ regime is equivalent to the
very small $h$ values. In this case, the energy gap of the  finite size system is basically representing 
the perturbative behavior. Thus, Eq.(\ref{gap}) is rewritten in the following form 
\begin{equation}
gap(N, h)=\frac{B}{N}+\frac{B}{N}f(x)\equiv \frac{B}{N}+ g(N) h^2,
\label{gap-pert}
\end{equation}
to a very good approximation.
Because, the first order 
perturbation correction is zero in the transverse field and the leading nonzero term is $h^2$. In this way 
we get, $g(N)=\frac{B f(x)}{N h^2}$. 
If the small $x$ behavior of  the scaling function defined by: $f(x)\sim x^{\phi_s}$ we
find that,
\begin{equation}
g(N)=B N^{(\phi_s -1)} h^{(\varepsilon \phi_s -2)}. 
\label{h2}
\end{equation}
Since $g(N)$ is a function of only $N$ we end up with 
\begin{equation}
\varepsilon \phi_s=2.
\label{phis}
\end{equation}
The multiplication of both sides of Eq.(\ref{gap-pert}) by $N$ leads to: $N gap(N, h)=B+B x^{2/\varepsilon}$.
This shows that in the small $x$ regime, $N gap(N, h)$ is a linear function of $x^{2/\varepsilon}$. 
This is in agreement with our data in Fig.(\ref{fig1}) where $\phi_s=1$ and according to 
Eq.(\ref{phis}), the value of $\varepsilon$ is found to be $\varepsilon=2$.

The function $g(N)$ in Eq.(\ref{gap-pert}) is actually the coefficient of the first nonzero correction
in the perturbation expansion for the energy gap of a finite chain,
\bea
gap(N, h)&=&gap(N, 0)+g_1(N) h^2+ \cdots +g_m(N) h^{2m} \nonumber \\
&=& \frac{B}{N}+g_1(N) h^2+ \cdots +g_m(N) h^{2m}
\label{pertexp}
\eea
where  $m$ is an integer. 
The effect of higher order terms can be neglected for $h\le 0.01$ to a very good approximation.
Now, let consider that the large $N$ behavior of $g_1(N)$ is,
\be
\lim_{N\rightarrow\infty} g_1(N) \simeq a_1 N^{\alpha}.
\label{e2}
\ee
We find that,
\be
gap(N, h)\simeq\frac{B}{N}\Big(1+b_1 N^{\alpha+1} h^2\Big),
\label{e3}
\ee
where $b_1=\frac{a_1}{B}=const$. We can write
Eq.(\ref{e3}) in terms of the scaling variable ($x=Nh^{\varepsilon}$),
\be
\frac{gap(N,h)}{\frac{B}{N}}\simeq 1+ N^{\alpha+1-\frac{2}{\varepsilon}}\;
x^{\frac{2}{\varepsilon}}.
\label{e4}
\ee
For large $N$ limit, Eq.(\ref{e4}) should be independent of $N$. This impose the following
relation,
\be
\alpha+1-\frac{2}{\varepsilon}=0.
\label{e5}
\ee
This equation defines the relation between $\alpha$ and 
$\varepsilon$,
\be
\varepsilon=\frac{2}{\alpha+1}.
\label{epa}
\ee
The above arguments propose to look for the large $N$ behavior of $g_1(N)$.
For this purpose, we have plotted in Fig.(\ref{fig2}) the  following expression versus $N$,
\be
g_1(N)\simeq\frac{gap(N, h)-gap(N, 0)}{h^2},
\ee
for  fixed values of  $h$ ($0.001\le h\le0.01$) and $\Delta=0.5, -0.5$. The results have
been plotted for different sizes, $N=12, 14, \dots, 24$ to derive the $\alpha$ exponent 
defined in Eq.(\ref{e2}). In Fig.(\ref{fig2}-a) we have considered the case of $\Delta=0.5$ and
found the best fit to our data for $\alpha=0.88\pm0.02$. Therefore, $\varepsilon=1.06\pm0.02$ which
 shows a very good agreement with Eq.(\ref{nu}), $\varepsilon_{FT}=\nu=12/11=1.09$.
Moreover, our data for different $h$ values
fall perfectly on each other which show that our results for  $g_1(N)$ is independent of $h$ as we have
expected. We have also plotted the results for  $\Delta=-0.5$ in Fig.(\ref{fig2}-b) and found $\alpha=0.55\pm0.02$.
Then, we obtain $\varepsilon=1.29\pm0.02$ which
can be compared with Eq.(\ref{mu}), $\varepsilon_{FT}=\mu=6/5=1.2$. This shows a slight deviation which
is the result of numerical computations and also the limitation on the size of system. 
Moreover, the magnitude of the energy gap is smaller for $\Delta=-0.5$ than the case of $\Delta=0.5$ which
implies less accuracy.
However, the results
for different $h$,  give once more a unique $g_1(N)$ in agreement with its definition to be independent of $h$.

We have extended our numerical computations to consider the other values of $\Delta$. The results 
have been presented in Table. I. We have listed  $\alpha$, the resulting $\varepsilon$ which
is obtained from Eq.(\ref{epa}) and the result of field theoretical approach ($\varepsilon_{FT}$) for different
values of $\Delta$. Our numerical  results show very good agreement with the exponents derived 
in the field theoretical approach\cite{dmitriev}.

%%%%%%%%%%%%%%%%%%%%%%%%%%%%%%%%%
\begin{table}[t]
\begin{center}
\label{table1} 
\caption{The $\alpha$ exponent 
defined in Eq.(\ref{e2}), the related  gap
exponent ($\varepsilon$) and the corresponding value  $\varepsilon_{FT}$ 
obtained by field theoretical approach for different anisotropy parameter
($\Delta$).}
\begin{tabular}
{|c|c|c|c|} \hline
$\Delta$  & $\alpha$ & $\varepsilon$ &  $\varepsilon_{FT}$\\
\hline
$0.70$ &$0.96$ &$1.02$   &$1.04$  \\
\hline
$0.50$ &$0.88$ &$1.06$   &$1.09$  \\
\hline
$0.25$ &$0.70$ &$1.17$   &$1.18$   \\
\hline
$0.0$ &$0.65$  &$1.21$    &$1.33$ \\
\hline
$-0.25$ &$0.60$ &$1.25$  &$1.26$     \\
\hline
$-0.50$ &$0.55$ &$1.29$  &$1.20$   \\
\hline
$-0.70$ &$0.59$  &$1.25$   &$1.14$ \\
\hline
\end{tabular}
\end{center}
\end{table}
%%%%%%%%%%%%%%%%%%%%%%%%%%%%%%%%%

%%%%%%%%%%%%%%%%%%%%%%%%%%%%%%%%%%%%%%%%%%%%%%%
%%%%%%%%%%%%%%%%%%%%%%%%%%%%%%%%%%%%%%%%%%%%%%%
%%%%%%%%%%%%%%%%%%%%%%%%%%%%%%%%%%%%%%%%%%%%%%%

\section{Discussion and summary}

We have studied the opening of excitation gap of the XXZ chain by
breaking the rotational symmetry. We have implemented the modified
Lanczos method to get the excited state energies at the same
accuracy of the ground state one. The convergence of this method
is fast for the low energy states but gets slower for higher
energy ones. The reason is related to the condensation  of states
in a fixed interval of energy by going to higher states. Thus the
method will be very sensitive to the initial parameter $E_t$
(Eq.(\ref{gros})). However, for the 10th lowest energy states we
got fast convergence. We have been limited to consider the maximum
$N=24$, because for the present model (Eq.(\ref{ham})) the total
$S^z$ does not commute with the Hamiltonian. Thus, we should
consider the full Hilbert space of $2^N$ in our computations.

We have tried to find the scaling of energy gap in the presence of the transverse field
by introducing the scaling function in Eq.(\ref{gaf1}). According to this approach,
the energy gap scales as $gap \sim h^{\varepsilon}$ where $\varepsilon$ defines
$x=N h^{\varepsilon}$, the scaling variable. The right scaling exponent gives 
a linear behavior of $N gap(N, h)$ versus $x$ for large $x$. To find the large $x$ behavior
we have faced with a serious problem. To see the scaling behavior we have to consider
very small values of $h$ to avoid the effect of level crossing between the excited state which
defines the gap and the upper states. The position of the first level crossing is roughly proportional
to the inverse of the chain length. Thus, we were not able to get the large $x$ behavior and being far from
the crossing point, at the same time. 
%We just found that the scaling function ($f(x)$) behaves 
%linearly in the small $x$ regime.

The limitation to very small $h$ values states that our results are representing the perturbative ones.
We have then led to get the scaling behavior by finding the divergence of $g_1(N) \sim N^{\alpha}$ 
in the thermodynamic limit ($N\rightarrow \infty$). The function $g_1(N)$ is the coefficient of the
leading term in the perturbation expansion (Eq.(\ref{pertexp})). Based on the formulation presented
in the previous section the gap exponent is related to the divergence of $g_1(N)$ by 
$\varepsilon=\frac{2}{\alpha+1}$. Our numerical results presented in Fig.(\ref{fig2}) show that
$g_1(N)$ is independent of $h$ and its divergence versus $N$ is defined by Eq.(\ref{e2}). 
Moreover, the gap exponent
which has been obtained by our numerics (listed in Table. I) is in very good agreement with
the results obtained by the field theoretical approach \cite{dmitriev}.

We might also find the gap exponent by finding the preceise location of the first 
level crossing between the second and third excited states which we call $h_1$.
The slope  of $log(h_1)$ versus $log(N)$ is $\frac{-1}{\varepsilon}$.
In the small $h$ regime where  Eq.(\ref{pertexp}) is approximated up to the  quadratic term, 
we can write two different expressions for the energy of the third and second excited states, namely
\bea
E_3(N, h)-E_0(N, h)=gap_3(N, h=0) + g^{(3)}_1(N) h^2, \nonumber \\
E_2(N, h)-E_0(N, h)=gap_2(N, h=0) + g^{(2)}_1(N) h^2.
\label{excited}
\eea
The subtraction of the two terms give the difference of $E_3-E_2$ which is
a function of the terms presented in the right sides of Eq.(\ref{excited}).
The zero field terms are proportional to $N^{-1}$ and the coefficients of the
quadratic terms obey Eq.(\ref{e2}). Thus, the scaling of $h_1$ in terms
of the lattice size is like, $h_1 \sim N^{-1/\varepsilon}$. However, the preceise
determination of $h_1$ defines the accuracy of this approach.

%%%%%%%%%%%%%%%%%%%%%%%%%%%%%%%%%%%%%%%%%%%%%%%%%
\acknowledgments
The authors would like to thank  K.-H. M\"utter, 
A. Fledderjohann, M. Karbach, I. Peschel, G. Japaridze and one of the referees 
for their valuable comments and discussions.
A.L. would like to thank the institute for advanced studies in basic
sciences where the initial part of this work was started.

\end{document}